\documentclass[pra,preprint,superscriptaddress,showpacs,preprintnumbers,amsmath,amssymb]{revtex4-1}

\usepackage{graphicx}
\usepackage{dcolumn}
\usepackage{bm}
\usepackage{ifpdf}
\usepackage{booktabs}

\begin{document}

\bibliographystyle{apsrev4-1} 


\title{Criterion of effective centre-of-mass method in Quantum Mechanics}
\author{Bo-Yuan Ning}
\affiliation{Institute of Modern Physics, Fudan University, Shanghai 200433, China}
\affiliation{Department of Optical Science and Engineering, Fudan University, Shanghai 200433, China}%
\author{Jun Zhuang}
\affiliation{Department of Optical Science and Engineering, Fudan University, Shanghai 200433, China}%
\author{Xi-Jing Ning}
\email{xjning@fudan.edu.cn}
\affiliation{Institute of Modern Physics, Fudan University, Shanghai 200433, China}
\affiliation{Applied Ion Beam Physics Laboratory, Key Laboratory of the Ministry of Education, Fudan University, Shanghai 200433, China}
\date{\today}

\begin{abstract}
In describing the motion of atoms and clusters, we face with choosing quantum mechanics or classical mechanics under different conditions.
In principle, there exist two criteria for this choice, but they do contradict in some cases though they are in agreement for other cases.
Actually, this problem is closely related with the effective centre-of-mass method, the underlying application of quantum mechanics.
It is shown that quantum mechanics must be selected for particle's motion when the de Broglie wave length of the mass centre is larger than the particle size, and in such case the effective centre-of-mass can be used in Quantum Mechanics.
In order to test this conclusion, an easy-manufactured experiment is suggested.
\end{abstract}


\maketitle

\section{Introduction\label{sec:1}}
As the development of nanoscale device, an important problem stands in the front of physicists that either classical mechanics (CM) or quantum mechanics (QM) is applied to the motion of nanoparticles.
It has been proved by lots of studies that the CM-based Monte Carol method or Molecular Dynamics is applied to the thermal motion of atoms at above room temperatures~\cite{zhang1997,Nissila}.
However, we have to adopt QM to deal with the interference of C$_{60}$ molecules~\cite{Arndt1999,Arndt2003} or the random vibration of a physical pendulum of monatomic carbon chain~\cite{Mazilova} although the mass of which are significantly larger than the mass of one atom.

In principle, there exists two well-established criteria between CM and QM.
The first one, put forward by Landau~\cite{Landau}, states that a particle will lose its quantum wave property and turn to be classical as the Planck's constant $\hbar\rightarrow0$ ($m\rightarrow\infty$).
The other is the uncertainty-principle-based measurement judgement, which asserts that QM must be used when the position uncertainty $\triangle x$ of the particle is on the same scale of the particle's de Broglie wavelength.
However, the two criteria do contradict with each other in some cases though they are equivalent in many situations.
As an example, the motion of a rigid ball, according to Landau's assertion, must perpetually obey classical Newton laws and can keep motionless as long as its mass $m\rightarrow\infty$;
on the contrary, the measurement judgement tells us that the ball, no matter how much its mass is, may transit in the eigenstates of the mass centre coordinate operator $\hat{\vec{R}}$, i.e., it may move faster than light once upon its mass centre is determined exactly ($\triangle \vec{R}\rightarrow0$).

In fact, this problem is closely related to the effective centre-of mass (ECM) method, which is underlying for both CM and QM framework because no realistic mass points have been identified experimentally.
The quantum explanation of C$_{60}$ interference phenomenon~\cite{Arndt1999} is the application of ECM, instead of considering the motions of elementary particles (electrons and nuclei) in the fullerene ball.
However, we cannot arbitrarily apply the ECM method in QM for any case though under some conditions Greiner has given a proof~\cite{Greiner}.
For instance, the Dirac wave equation for single particle without inner structure correctly predicts the spin magnetic moment of an electron, but it fails to give the correct value of neutron~\cite{Stern}.
This is understandable because a neutron consists of three smaller particles (quarks).
Thus, it is needed to find out a criterion for correctly applying the ECM method in QM.

In this work, we theoretically analyze the two criteria for distinguishing QM from CM and propose a specifical judgement for validation of ECM in QM.
In order to test the conclusions, an easily manufactured experiment is suggested to observe the motion of a physical pendulum made up of a single-walled carbon nanotube (CNT), which will behave quantum oscillation without period as long as its arm length is on the order of micrometers and the mass is less than $10^{-19}$Kg.

\section{Criterion for CM and ECM method in QM\label{sec:2}}
When solving the motion of real objects, such as C$_{60}$, carbon tubes or atomic chain, we have to select between the Newton's equation and the Schr\"{o}dinger one, because the latter is much more complicated.
From the viewpoint of Landau~\cite{Landau}, the wave function $\Psi$ of a physical system can be written as
\begin{equation}
\label{eq1}
\Psi=ae^{iS_c/\hbar}.
\end{equation}
By an analogy with the principle of least action utilized in optics, Landau concluded that QM will turn to CM as long as the Planck's constant $\hbar\rightarrow0$ .
A way to make the constant $\hbar$ disappear is to let the mass of the particle tends to infinite, $m\rightarrow\infty$.

Another judgement is based on the Heisenberg's uncertainty principle,
\begin{equation}
\label{eq2}
\triangle x\ge \frac{\hbar}{\triangle p}.
\end{equation}
Since the $\hbar/\triangle p$ is on the order of the de Broglie wave length of the particle, $\lambda$, the particle will behave quantum motion if its moving space is limited less than $\lambda$.
Although the two criteria are consistent in many situations, they do obviously contradict with each other in some situations, which can be seen in the entropy calculation of a system consisting of $N$ ideal particles~\cite{Morandi}.
Classical statistics shows that the entropy $S$ of such a system is
\begin{equation}
\label{eq3}
S=\frac{5}{2}Nk_B+3Nk_B\ln(\frac{\bar{d}}{\lambda_{T}}),
\end{equation}
where $k_B$ and $\bar{d}$ are Boltzmann constant and the mean free path of the ideal particles respectively, and thermal wave length $\lambda_{T}=h/\sqrt{2\pi m k_B T}$ with $T$ the temperature.
For a given $\bar{d}$, there exists a critical temperature point $T^*=h^2/(2\pi m k_B \bar{d}^2)$ at which $S$ would vanish.
According to Landau's judgement, the system will follow CM rules if $m$ is large enough and Eq.~(\ref{eq3}) will be valid, and for the finite $m$ as the common sense, Eq.~(\ref{eq3}) predicts a temperature interval $0<T<T^*$ in which $S$ is less than zero and tends to be negative infinite as $T\rightarrow0$.
This wrong result can be removed off by the other criteria for the case of $T<T^*$: The de Broglie wave length $\lambda_T$ of the particles is far longer than $\bar{d}$ when $T<T^*$, and Eq.~(\ref{eq3}) should be replaced with quantum statistics because CM no longer holds.

The above example indicates that the measurement judgement is more reasonable, which can be further enhanced by path integral theory~\cite{Kleinert}.
The propagator of a free particle of mass $m$ reads
\begin{equation}
\label{eq4}
<q',t'|q,t> =  \sqrt{\frac{m}{2\pi i \hbar (t'-t)}}\exp \left[\frac{im}{2\hbar}\frac{(q'-q)^2}{t'-t}\right],
\end{equation}
which states that one particle initially ($t$) located at spatial point $q$ may be observed anywhere with the same probability.
Clearly, this quantum phenomenon always occurs no matter how large the mass is.

Now, consider a macroscopic pendulum with periodic oscillation, which can be easily understood by Landau's judgment because the mass of the pendulum ball is large enough. From the measurement judgement, however, if the ECM method in QM is still valid for this pendulum, then the ball would never do periodic motion once we exactly measure the position of the mass center via, e.g., a weak laser beam.
This fact implies that the validation of the ECM method in QM needs some conditions.

\begin{figure}
\centering
\includegraphics[width=3.25in,height=1in]{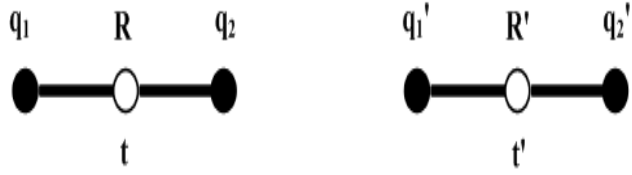}
\caption{The mass centre of a particle composed of two mass points distant by $2d$ at time $t$ moves from spatial point $R$ to $R'$ at time $t'$ with the coordinates of the mass points changed from $q_{1,2}=R\pm d$ to $q_{1,2}'=R'\pm d'$.}
\label{fig1}
\end{figure}

For obtaining the conditions, we consider a particle composed of two mass points each with mass $m$ (Fig.~\ref{fig1}).
If the interaction between the mass points is weak enough to be neglected, then the Feynman kernel for one-dimension motion is
\begin{equation}
\label{eq5}
<q_1', q_2', t'|q_1, q_2, t>=C\exp\left[\frac{im(q_1'-q_1)^2}{2\hbar(t'-t)}]\exp[\frac{im(q_2'-q_2)^2}{2\hbar(t'-t)}\right].
\end{equation}
Supposing the distance between the mass points at time $t$ and $t'$ is $2d$ and $2d'$, Eq.~(\ref{eq5}) can be rewritten as
\begin{equation}
\label{eq6}
<q_1', q_2', t'|q_1, q_2, t>=C\exp\left[\frac{i2m[(R'-R)+D]^2}{2\hbar(t'-t)}\right],
\end{equation}
where $D=d'-d$ approximates to the particle size.
If the ECM method is valid for the particle, then, according to Eq.~(\ref{eq4}), the Feynman kernel should be
\begin{equation}
\label{eq7}
<R', t'| R, t>=C\exp\left[\frac{i2m(R'-R)^2}{2\hbar(t'-t)}\right].
\end{equation}
Comparing Eqs.~(\ref{eq7}) with (~\ref{eq6}), we get the condition for ECM in QM, $R'-R=\triangle R\gg D$.
Certainly, the validation of ECM does not mean that the particle must behave quantum motion.
Taking the measurement judgement for QM ($\triangle R\sim \lambda$) into account, we conclude that when
\begin{equation}
\label{eq8}
\lambda \gg D,
\end{equation}
the particle's motion must be described by QM and can be treated by ECM method.

For a macro-size ball of a common pendulum, the de Broglie wave length is much smaller than its diameter, and therefore it displays classical periodical oscillation.
If the ball size is reduced approaching to zero without changing its mass, it will, according to Eq.~(\ref{eq8}), do quantum random motion without any period.
For the thermal motion of atoms at above room temperatures, the $\lambda_M$ is about 0.1{\AA}, which is much smaller than atom's size 1{\AA}, and CM-based molecular dynamics can be used to simulate the motion.
It should be stressed that the condition (Eq.~(\ref{eq8})) may be valid in one or two spatial freedoms even though it is not in the other freedoms.
As an example, the motion of the mass centre of a carbon atomic chain handed at one end satisfies the condition in the tangential direction (Fig.~\ref{fig2}), but violates the condition in the radius direction~\cite{Mazilova}.
Clearly, for a given object, QM may be used in one freedom with CM applied to other freedoms simultaneously.

\section{Quantum behaviors of a physical pendulum\label{sec:3}}
\begin{figure}
\centering
\includegraphics[width=2in,height=2.5in]{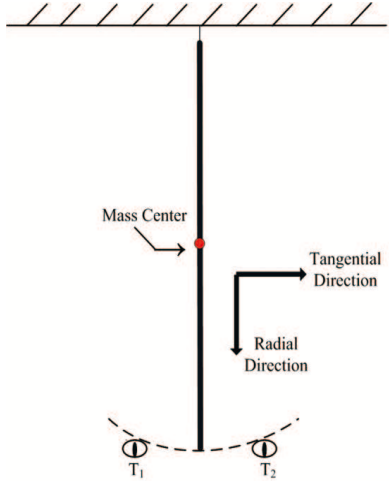}
\caption{Schematics of a physical pendulum. A single-walled CNT suspended by a short monatomic chain serving as a physical pendulum, observed via two microprobe detectors, T$_1$ and T$_2$.}
\label{fig2}
\end{figure}

\begin{figure}
\centering
\includegraphics[width=3.25in]{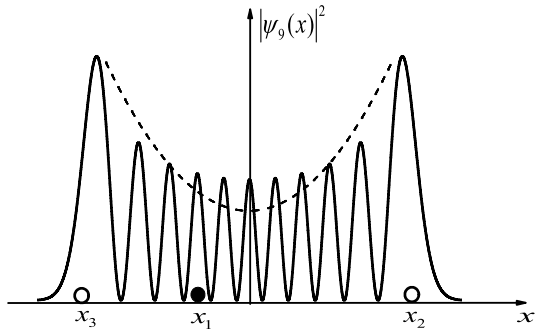}
\caption{The motion of a physical pendulum predicted by QM with ECM applied. For the mass centre initially located at $x_1$, it may randomly transit to $x_2$ or $x_3$.}
\label{fig3}
\end{figure}

In order to test the criterion Eq.~(\ref{eq8}), we design an easy-manufactured experiment on a physical pendulum.
What we expect is that the motion of the pendulum will change from periodic oscillation to random jumping when the condition is satisfied.
With the ECM method, the motion of a physical pendulum of mass $M$, which may be a macro size pendulum or a carbon nanotube (Fig.~\ref{fig2}), can be treated as 1-dimensional motion of mass point under a potential $V(x)=\frac{1}{2}M\omega^2x^2$, where $\omega=\sqrt{\frac{3g}{2l}}$ is the angular frequency for arm length $l$.
In CM regime, the ECM motion is the harmonic oscillation with a period of $\frac{2\pi}{\omega}$.
In QM framework with ECM method applied, the pendulum will be in one of its ECM's discrete energy eigenstates, $\psi_n(x)$, with eigenvalue $E_n=(\frac{1}{2}+n)\hbar\omega$, where
\begin{equation}
\label{eq9}
\psi_n(x)=\frac{M\omega}{\pi\hbar}^{\frac{1}{4}}(2^nn!)^{-\frac{1}{2}}e^{-\frac{M\omega}{2\hbar}x^2}H_n\left(\sqrt{\frac{M\omega}{\hbar}}x\right) ~~~ n=0,1,2\ldots
\end{equation}
with $H_n(\xi)$ the Hermite polynomial.
According to QM explanation, the ECM will instantaneously jump randomly to any spatial position $x$ with the probability $|\psi_n(x)|^2$ even if its mass tends to be infinity (Fig.~\ref{fig3}), which is very different from the classical periodical oscillation.
We never observed such QM transition for common macro-size pendulum just because the condition Eq.~(\ref{eq8}) is not fulfilled, and indeed, Mazilova group~\cite{Mazilova} did observe non-periodic spatial transition of a monatomic carbon chain lying in the ground state and the first exited one, as $|\psi_{n}(x)|^2$ predicts.

\begin{table}
\caption{\label{tab1}Conditions for QM motion of a single-walled CNT in the 5th eigenstate.}
\begin{ruledtabular}
\begin{tabular}{cccccc}
 l(nm) & M(10$^{-21}$Kg) & $\lambda_{min}(nm)$ & Amplitude(nm) & Angle($\textordmasculine$) & Period(s)\\
 \colrule
 1.5$\times$10$^{7}$ & 20100 & 0.25 & 3.9    & 1.41$\times$10$^{-5}$ & 0.20\\
 5.5$\times$10$^{4}$ & 184   & 1.99   & 9.7   & 9.58$\times$10$^{-3}$ & 0.01\\
 9$\times$10$^{2}$   & 6.04  & 3.95   & 18.8   & 1.13                  & 1.55$\times$10$^{-3}$\\
 7$\times$10$^{1}$   & 0.70  & 6.12   & 28.6   & 22.1                  & 4.33$\times$10$^{-4}$\\
 1.3$\times$10$^{1}$ & 0.17  & 8.16   & 37.4   & 156                   & 1.87$\times$10$^{-4}$\\
\end{tabular}
\end{ruledtabular}
\end{table}

For a common macro-size physical pendulum of mass $M$ (Fig.~\ref{fig2}), the minimum de Broglie wave length $\lambda_{min}$ along tangential direction follows as
\begin{equation}
\label{eq10}
\lambda_{min}=2\pi\sqrt{\frac{\hbar}{(2n+1)M}\sqrt{\frac{2l}{3g}}}.
\end{equation}
According to Eq.~(\ref{eq8}), if the diameter of the tube $D$ is much smaller than $\lambda_{min}$, the physical pendulum will display quantum transition.
For the monatomic carbon chain in Mazilova's experiment~\cite{Mazilova}, the corresponding $\lambda_{min}\simeq180nm$, which is greatly larger the chain diameter 0.2nm and therefore quantum spatial transition must take place.

A single-walled CNT is a good candidate for a quantum physical pendulum as shown in Fig.~\ref{fig2}.
Currently, it is available to obtain macro-size single-walled CNTs of millimeters in length with about $0.4nm$ diameters in common laboratory~\cite{Qin,Wang2000}.
Although it seems impossible for a CNT to display quantum movement because of its macroscopic body, the QM motion along the tangential direction only requires the $\lambda_{min}$ larger than the tube's diameter ($0.4nm$).
It should be noted that the quantum spatial transition covers larger distances as $n$ increases, which benefits the observation of the quantum motion, but the $\lambda_{min}$ is on the contrary trend for a given pendulum length.
So, in the numerical estimation of the condition for quantum motion, we let the CNT lying in its 5th eigenstate ($\psi_5(x)$) and attained the specific conditions shown in Tab.~\ref{tab1}, where the last column gives the corresponding classical oscillation periods.

As shown in table.~\ref{tab1}, the physical pendulum is quite suitable for detecting quantum transition via two microprobes placed near to the free end of the tube.
For example, the single-walled CNT as long as $1.5cm$ should display classical oscillation of 5Hz (see the data in the first row of Tab.~\ref{tab1}) because the condition (Eq.~(\ref{eq8})) is not fullfilled.
As the tube gets shorter and shorter, QM random vibration should occur.
Considering larger motion range of the free end is benefited to the arrangement of the two detection probes, we could use a tube of 55$\mu m$ in length (see the data in the second row of table.~\ref{tab1}) with a vibration range of about $19nm$.
In order to observe the classical pendulum turning to the quantum one, a longer CNT may be used first to show the classical periodical oscillation, and then cut a piece of the tube until it satisfies the condition in table.~\ref{tab1} to display quantum movement.

\section{Conclusion\label{sec:4}}
In conclusion, the ECM method should be conditionally applied in QM, and on the premise of the validation of ECM, it is shown that the uncertainty-principle-based measurement judgement is more reasonable than the one provided by Landau.
This conclusion is supported by the experiments on the interference of C$_{60}$ molecules and the quantum motion of
carbon monatomic chain, and would be further confirmed by the physical pendulum suggested above.
For general application of QM, we must compare the de Broglie wave length with the particle's size.
As an example, the de Broglie wave length of carbon atoms in condensed matter at above room temperatures is much smaller than the atom's size, and CM can be used for describing the atom's motion. However, when the environmental temperature approaches to absolute zero degree, we have to adopt QM for the atomic motion and ECM method can be employed because the de Broglie wave length is much larger than the size of the atoms.


\acknowledgments  This work is supported by the Innovation Program of the Shanghai Municipal Education Commission under Grant No. 10ZZ02, Key Discipline Innovative Training Program of Fudan University and National Natural Science Foundation of China Grant No. 11074042.

%

\end{document}